\documentclass[12pt]{article}

\usepackage{amsmath,amssymb}
\usepackage{amsfonts}

\newcommand{\tr}{\mathrm{Tr}}
\newcommand{\extd}{{\rm d}}

\newcommand{\tJ}{\tilde{J}}

\def\be{ \begin{equation}}
\def\ee{\end{equation}}
\def\bes{\begin{eqnarray}}
\def\ees{\end{eqnarray}}

\begin{document}

\renewcommand{\thefootnote}{\fnsymbol{footnote}}
\centerline{\Large \bf On  pure Yang-Mills theory in $3+1$ dimensions:}

\centerline{\Large \bf Hamiltonian, vacuum and gauge invariant variables}
\vskip 0.75 cm

\centerline{{\bf 
Laurent Freidel,${}^{1,2}$\footnote{lfreidel@perimeterinstitute.ca}
}}
\vskip .5cm
\centerline{${}^1$\it Perimeter Institute for Theoretical Physics,}
\centerline{\it 31 Caroline St. N. Waterloo, N2L 2Y5, ON, Canada. }
\vskip .5cm
\centerline{${}^2$\it Laboratoire de Physique, Ecole Normale Sup{\'e}rieure de Lyon}
\centerline{\it 46 All{\'e}e d'Italie, 69364 Lyon, Cedex 07, France.}

\setcounter{footnote}{0}
\renewcommand{\thefootnote}{\arabic{footnote}}

\begin{abstract}
In this work we discuss an analytic approach towards the solution of
pure Yang-Mills theory in $3+1$ dimensional spacetime which 
strongly suggests that the recent strategy already applied to pure Yang-Mills theory in $2+1$ can be extended to $3+1$ dimensions.  
We show that the local gauge invariant variables introduced by Bars gives a natural 
generalisation to any dimension of the formalism of Karabali and Nair which recently led to 
a new understanding of the physics of QCD in dimension $2+1$.
After discussing the kinematics of these variables, we compute the jacobian between the Yang-Mills and Bars variables
and propose a regularization procedure which preserves a generalisation of holomorphic invariance.
We discuss the construction of the QCD hamiltonian properly regularized and compute the behavior 
of the vacuum wave functional both at weak and strong coupling.
We argue that this formalism allows the developpement of a strong coupling expansion in the continuum
by computing the first local eigenstate of the kinetic part  of Yang-Mills hamiltonian.

\end{abstract}

\section{Introduction}
In a remarkable series of papers Karabali and Nair \cite{KN} and Karabali, Kim and Nair \cite{KKN} (see \cite{Schulz} for an introductory review)
have developed a novel and powerful method in order to adress the confinement problem of pure Yang-Mills 
in dimension 2+1 (I will refer to this work as KKN for short). 
Their approach, inspired by the work of Feynman \cite{Feynman} on the Schr\"odinger representation of QCD,
 has led to an analytic determination of the string tension and the possibility to devise a strong coupling expansion in the continuum
and evidence for a mass gap. The main idea behind this work is to formulate the Yang-Mills hamiltonian in
 terms of {\it local gauge invariant} variables. 
This parametrization heavily uses the power of the complex structure which is available in 2D.

This work has been followed by by a deeper investigation of the vacuum wave functional in the Large N limit by Leigh, Minic and Yelnikov \cite{prl,long}.
Basing their analysis  on a quadratic ansatz for the wave functional, these authors have investigated the structure of the Schr\"odinger equation
and the action of the kinetic term on a restricted class of operators. From this study  they have proposed 
an interpolating kernel capturing the behavior of the vacuum wave function both in the infrared and the ultraviolet.
This kernel, expressed simply as a ratio of Bessel functions, exhibits an infinite set of resonances arguing in favor of a constituent picture 
for glueballs which becomes apparent in the KKN formalism. This have led them to a theoretical prediction of the 
glueball spectra simply related to the zeros of Bessel function 
which is in striking agreement with the large N lattice data \cite{Teper}.

This series of work open many new questions and avenues in the study of QCD, first in 2+1 dimensions 
where among others things one can wonder about the possibility of exploring along these lines the mesons spectra, the inclusion of a non zero temperature and 
where a deeper understanding of the renormalization group in this context is needed.
But one of the most pressing issue is to know wether such a methodology can be extended and applied to the case of pure Yang-Mills in dimension 3+1.
The main message of the present work is to show that this is indeed the case.

The first obstacle to overcome lies in the fact that the KKN formalism uses the full strength of complex analysis which is available in 
2 space dimensions but not in 3D. The key and simple insight obtained in collaboration with R. Leigh and D. Minic \cite{jointpaper}
is the fact that the KKN gauge invariant variables are related to lines integral from infinity to a point $x$ and admit a natural generalization in any dimension which
was first proposed by I. Bars \cite{bars} some time ago and baptized  `corner variables'.
 
In this work we explore the consequences of this insight and develop a new set of techniques allowing us to work in terms of the Bars variables and extract 
a first set of physical information from this formalism.
In section \ref{Barsvar} we explore the kinematics of Bars variables and reestablish some important formulae 
relating them to the usual Yang-Mills variables, we also identify the key notion of an `holomorphic symmetry'.
In section \ref{det} we present one of our first main result which is the computation of the jacobian of the transformation 
between Yang-Mills and Bars variables. This computation is possible once we devise a natural regularization scheme
 which preserves holomorphic symmetry.
The jacobian is then shown to be  trivial in the case of real coordinates and given in terms of a generalization of the 
WZW action in the case of semi-complex coordinates.
This result implies that the trivial wave functional is normalizable and hence the theory is amenable to a strong coupling expansion.
This  strongly resonates with the results obtained in lattice QCD but in the context of a continuum formulation.

In section \ref{ham} we present the construction of the regularized hamiltonian in Bars variables and as a check give the construction 
of the wave functional in the ultraviolet regime.

In section \ref{vac} we show that the regularized kinetic term acts diagonally on the potential term which is our second main  result. In contrast with 
2+1 D the eigenvalue, having a dimension of a mass, is cutoff dependent. 
This allows us to construct the vacuum wave function at first order in a strong coupling expansion and using an argument 
of dimensional reduction we show how 
this  determines the value of the string tension in this regime.
We conclude on the open issues, one of the main one being the problem of the continuum limit.

In a joint companion paper \cite{jointpaper} we explore the physical consequences of the results obtained here
and strongly argue in favor of a deep similarity between the 2+1 and 3+1 dimensional case.


\bigskip
\section{The kinematics of Bars Corner variables}\label{Barsvar} 
We denote by $A_{i}$, $i=1,2,\cdots D$ the spatial components of the Yang-Mills connection. 
These are taken to be anti-hermitian and our main interest is $D=3$, but we keep it free in the first part of the 
paper. One of the key ingredient of Karabali Kim and Nair (KKN for short) formalism is a parametrization of 
the configuration space of $2d$ Yang-Mills in terms of local gauge invariant variables which uses heavily at first sight 
the available complex structure of 2 dimension. 
In fact \cite{jointpaper}  the KKN gauge invariant variables can be understood as Wilson lines integral from infinity 
to a point $x$ and admit a natural generalization in any dimension which
was proposed by I. Bars \cite{bars}. We denote them $M_{i}(x)$, they satisfy the defining equation
\be\label{defm}
A_{i}=- \partial_{i}M_{i} M_{i}^{{-1}}.
\ee
No summation over repeated indices is assumed here and after.
A solution is given by
\begin{equation} 
M_j(x) = \overleftarrow{exp} [-\int_{-\infty}^x A]
\end{equation} 
where the integral is a straight spatial contour for fixed $x^i$ for $i\neq j$ and $\overleftarrow{\exp}$ is the path ordered exponential, 
explicitly
\be\label{ordexp}
M_{1}(x) = 1 +\sum_{n=1}^\infty (-1)^n \int_{x_{1}> t_{1}> \cdots >t_{n}} \!\!\!\!\!A_{1}(t_{1},x_{2},x_{3}) \cdots A_{1}(t_{n},x_{2},x_{3}) 
\extd t_{1}\cdots \extd t_{n}.
\ee
 Gauge transformations 
\be
A_{i}\to A_{i}^{g}= gA_{i}g^{-1}+ g\partial_{i}g^{-1} 
\ee act on $M$'s as 
\begin{equation}
 M_i\to g M_i.
\end{equation}
One can then define gauge invariant local variables 
\begin{equation}
H_{ij}=M_i^{-1}M_j.
\end{equation}
Note that $H_{jj}=1$ and $H_{ji}=H_{ij}^{-1}$ --this just means traversing the corner in the opposite direction is precisely the inverse element. 
The $H_{ij}$ are {\it unitary} in a real coordinate basis. There is also a constraint (here written for $D=3$)
\begin{equation}
H_{ij}H_{ji}= 1, \quad  H_{ij}H_{jk}H_{ki}=1.
\end{equation}

The translation to 2+1 KN variables is obtained if one uses a complex basis 
\begin{equation} M_z=M,\ \ \ \ \ M_{\bar z}=(M^{\dagger})^{-1},\quad  H\equiv H_{\bar{z}z}= M^{\dagger}M
\end{equation}
Note that if we had used a real coordinate basis, then we would have had a pair of (unrelated) {\it unitary} matrices $M_1$, $M_2$.
In 3+1 there is no complex structure and we will work in a real basis where $M_{i}$ are unitary matrices.
One could also choose to work in a  ``semi-complex'' coordinate basis $\{ u,z,\bar z\}$ for $D=3$, then we could parameterize this as
\begin{eqnarray}
H_{uz}&=&H\\
H_{\bar z u}&=&H^\dagger\\
H_{\bar z z} &=& H^\dagger H 
\end{eqnarray}
For example, one could use the notation $M_z=M$, $M_{\bar z}=(M^{\dagger})^{-1}$, $M_u^\dagger M_u=1$, with $H=M_u^{\dagger}M$. 
The constraint takes the form $H_{\bar zu}H_{uz}H_{z\bar z}=1$.
In other words, there is in $D=3$ in the semi-complex coordinate basis a {\it complex} $H$-field (compared to a Hermitian field in $D=2$); thus there are 
twice as many degrees of freedom.

A very important feature of this parametrization is a generalization to any dimension of holomorphic invariance, and even if there is no complex structure in dimension $3$
we will refer to this symmetry as holomorphic symmetry.
 This extra symmetry acts as
\begin{equation} 
M_i\mapsto M_{i}h_i^{-1}(x^j),\ \ \ \ j\neq i\end{equation}
The condition $j\neq i$ on the function $h_i$ is the analogue of holomorphy.
 This leaves the gauge fields invariant, and one finds
\begin{equation} H_{ij}\mapsto h_iH_{ij}h_j^{-1}\end{equation}
In the complex basis, we would have $M_u\to M_{u}h_u^{-1}(z,\bar z)$, $M^\dagger\to h (u, z)M^{\dagger}$, $M\to Mh^{\dagger}(u,\bar{z})$ and so
$H\to h_u(z,\bar z)H h^\dagger (u,\bar z)$, $H^\dagger\to h(u,z)H^\dagger h_u^{-1}(z,\bar z)$.
Note that one can introduce currents
\begin{equation} 
J_{ij}=(\partial_j H_{ij}) H^{-1}_{ij},
\end{equation}
(in 2+1, $J\sim J_{\bar zz}$ and $J^\dagger\sim -J_{z\bar z}$). 
The $J_{ij}$ transform as (`holomorphic') connections
\begin{equation} J_{ij}\mapsto h_i J_{ij}h_i^{-1}+\partial_j h_i\ h_i^{-1}\end{equation}
(This is not a typo -- it only depends on $h_i$.)

In the real coordinate basis, it appears that there are six currents that are apparently distinct. 
However, there is a 'reality' condition on their derivatives of the form
\begin{equation}\label{dJ}
\partial_i J_{ij}=-H_{ij}(\partial_j J_{ji})H_{ij}^{-1}
\end{equation}
(in $D=2$, this reads $\bar\partial J= H(\partial J^\dagger) H^{-1}$.) By defining $\bar J_{ij}=-H_{ij}J_{ji}H_{ij}^{-1}$, we may rewrite this as
\begin{equation}
\partial_i J_{ij}=\partial_j\bar J_{ij}-[J_{ij},\bar J_{ij}]
\end{equation}
and so there are covariant derivatives $D_{ij}=\partial_j-J_{ij}$.
These currents are related to the magnetic field 
 $F_{ij}= \partial_{i}A_{j}-\partial_{j}A_{i}+ [A_{i},A_{j}]$
by 
\be 
\partial_i J_{ij}= - M_{i}^{{-1}}F_{ij}M_{i}.
\ee
Note that the covariant derivatives can be written in terms of the usual derivative and $M_{i}$ as 
\be\label{covd}
\nabla_{i}=\partial_{i}+ A_{i}= M_{i}\partial_{i} M_{i}^{-1}.
\ee
 
Let us finally note that Wilson loops observables can be expressed entirely in terms of the currents:
Lets consider $C$ a closed curve in the $12$ plane. The Wilson loop observable
 in the representation $R$ can be expressed as
\be
W_{R}(C) = \tr_{R}\overleftarrow{e}^{-\oint_{C} A_{i}d {x_{i}}} = \tr_{R}\overleftarrow{e}^{\oint_{C} J_{12}(x(t)) \extd x_{2}}.
\ee
This is easily obtained once we write the connection as a gauge transformation of the current
\bes
A_{1}dx_{1}+A_{2}dx_{2}&=& -\extd x_{1}\partial_{1}M_{1}M_{1}^{-1}- \extd x_{2}\partial_{2}M_{2}M_{2}^{-1}\\
&=& -(\extd x_{1}\partial_{1}M_{1}M_{1}^{-1} +
\extd x_{2} (M_{1}J_{12}M_{1}^{-1} -\partial_{2}M_{1}M_{1}^{-1}))
\ees
If one works out what this correspondence imply for a rectangular plaquette $C$ in the plane $y_{3}=cste$ 
with corners $(0,0)  (y_{1},0 ) (0,y_{2}) (y_{1},y_{2})$ oriented clockwise one simply get 
\bes
W_{R}(y) \nonumber
 &=& \tr_{R}\left( \overrightarrow{e}^{-\int_{0}^{y_{2}}J_{12}(y_{1},x_{2})dx_{2}}\,
\overleftarrow{e}^{\int_{0}^{y_{2}}J_{12}(0,x_{2})dx_{2}}\right)\\
&=& \tr_{R}\left(  H_{12}(y_{1},0) H_{12}^{-1}(y_{1},y_{2}) H_{12}(0,y_{2})H_{12}^{-1}(0,0) \right)
\label{Willoo}
\ees
In order to continue and work out some formulae necessary for the computation of the hamiltonian in these variables lets introduce some notations and conventions.
The connection is expanded in terms of {\it anti-hermitian }generators $T_{a}$ (in order to avoid unnecessary factors of $i$), $A_{i}= A_{i}^{a} T_{a}$
satisfying the algebra $[T_{a}, T_{b}]= f_{ab}{}^{c}T_{c}$, we also denote by $\tr$ the trace in the vectorial representation, so that $\tr(1)=N$.
by  $- 2 \tr(T_{a}T_{b})= \delta_{ab}$
(the minus sign is because of the antihermiticity and the 2 is the standard convention).  The Yang-Mills action is taken to be $S_{YM}= \frac1{2g^2} \int\tr(F_{\mu\nu}F_{\mu\nu})$
and the hamiltonian in the Yang-Mills variables is 
\be {H= \sum_{i,a}\int -\frac{g^2}{2} \left(\frac{\delta}{\delta A_{i}^{a}}\right)^2 + \frac{1}{2g^2} (F_{i}^{a})^2}\ee with $F_{i}^{a}=\frac12 \epsilon_{ijk}F_{jk}^{a}$.
This should be supplemented by the gauss law constraint $\nabla_{i}^{ab}\frac{\delta}{\delta A_{i}^{b}} =0$.

In the adjoint representation the generators are given by $(T^{a})_{bc}= -f_{abc}$, where index are raised or lowered with the metric $\delta_{ab}$ and the trace in the adjoint is 
denoted by $\tr_{ad}$. A group element $M$ is represented in the adjoint by $M_{ab}= -2\tr(T_{a}MT_{b}M^{-1})$, clearly we have 
$(M^{-1})_{ab}= M_{ba}$ and also $MT_{b}M^{-1}= T_{a} M^{a}{}_{b}$.

We denote $G_{i}(x,y)$ the inverse of $\partial_{i}$, $\partial_{i}^x G_{i}(x,y)= \delta(x,y)$ (no summation on repeated indices).
Since $\partial_{i}$  admits zero modes, its inverse is not uniquely defined and we will work with the explicit choice
$G_{1}(x)=  \theta(x_{1})\delta({x_{2}})\delta({x_{3}})$ where $\theta$ is the heaviside  function and $G_{i}(x,y)\equiv G_{i}(x-y)$ (which is not antisymmetric).
This choice is not arbitrary, it is the unique 
 choice consistent with the definition of the variables $M_{i}$ as an ordered exponential. Indeed we can rewrite (\ref{ordexp}) in terms of the propagator 
as
\be \label{ordprop}
M_{i}(x) = \sum_{n}(-1)^n \int \extd y (G_{i}A_{i})^n(x,y),
\ee
where $(G_{i}A_{i})^2(x,y)= \int \extd z G_{i}(x,z)A_{i}(z)G_{i}(z,y)A_{i}(y) $ etc...

One can first compute the derivative of $M_{j}$ with respect to $A_{i}$. Starting from
$$ A_{i}^{a}= 2\tr(T^{a}\partial_{i}M_{i}M_{i}^{-1}),$$ one obtains  the relation 
$$\delta A_{i}^{a}= -(M_{i})^{ab} \partial_{i}(M_{i}^{-1}\delta M_{i})^{b}$$
with the obvious notation $M_{i}^{-1}\delta M_{i} = (M_{i}^{-1}\delta M_{i})^{b}T_{b}$. Inverting this relation gives 
\be \label{MA1}
\frac{(M_{j}^{-1}\delta M_{j}(y))^{b}}{\delta A_{i}^{a}(x)} = -\delta^{i}_{j}  G_{i}(y,x)(M_{i}^{-1}(x))^b{}_{a}= \delta_{i}^j(M_{i}G_{i}^t)(x,y)_{a}{}^b.
\ee
where we introduced
the notation 
\be
G^t(x,y) \equiv -G(y,x).
\ee
$t$ stands for transpose  and  the minus sign insures that $\partial_{i}^{x} G^t(x,y) = \delta(x,y)$.
In other words  we can express $A_{i}$ derivatives in terms of right derivative on the group
$$[P^{i}_{a}(x), M_{j}(y)] \equiv \delta^{j}_{i} M_{i}(y) T_{a} \delta(x,y)$$ as 
\be \label{MA2}
\frac{\delta}{\delta A_{i}^{a}(x)} = (M_{i}(x))_{a}{}^{b}\int \extd y \,G_{i}^t(x,y)P^{i}_{b}(y) \equiv (M_{i}G_{i}^tP^{i})_{a}(x).
\ee
The next step is to express $A$ derivatives in terms of the currents, from their definitions one has 
\be \nonumber
J_{ij}= \partial_{j}H_{ij}H_{ij}^{-1}= - M_{i}^{-1}\left(A_{j} + \partial_{j}M_{i}M_{i}^{-1}\right)M_{i}
=- M_{i}^{-1}A_{j}M_{i} -  M_{i}^{-1}\partial_{j}M_{i},
\ee
taking its variation one gets
\be
\delta J_{ij} = - M_{i}^{-1}\delta A_{j}M_{i} - [D_{ij}, M_{i}^{-1}\delta M_{i}]
\ee
with $D_{ij}=\partial_{j}-J_{ij}$.
If one takes (\ref{MA1}) into account this reads 
\be
\delta J_{ij} = - M_{i}^{-1}\delta A_{j}M_{i} + [D_{ij},G_{i}M_{i}^{-1}\delta A_{i}M_{i} ],
\ee 
from which we  get  
\bes
\frac{\delta J_{ji}^b(y)}{\delta A_{i}^{a}(x)}&=& -(M_{j}(x))_{a}{}^{b}\delta(x,y), \\
\frac{\delta J_{ij}^b(y)}{\delta A_{i}^{a}(x)}&=& ( D_{ij}^y)^{b}{}_{c} G_{i}(y,x)( M_{i}^{-1}(x))^{c}{}_{a}\\
\quad \frac{\delta J_{jk}^b(y)}{\delta A_{i}^{a}(x)}&=&  0, \, \,\, i\neq j,k
\ees
where $$ D_{ij}^{ab} = \partial_{j} \delta^{ab} + J_{ij}^{c}f_c{}^{ab}$$ (note that $(T_{c})_{ab}= -f_{cab}$).

So far everything is similar to 2D. 
Before going on, it is useful 
to dwell further on the kinematical structure and write down more precisely the structure of 
phase space in this new variables. 
In order to do so we need some notations:  one introduces the momenta variables which replace the electric field generators in our variables
and the corresponding operator smeared with a Lie algebra valued 1-form field $\Phi_{i}(x)= \Phi_{i}^{a}(x)T_{a}$
\be
\Pi_{a}^{i}(x) \equiv (G_{i}^{t}P^{i})_{a}(x), \quad \Pi(\Phi) \equiv \int \extd x \, \Phi^{a}_{i}(x) \Pi_{a}^{i}(x).
\ee
Obviously, 
\be 
\Pi(\partial \Phi) = - P(\Phi), \quad \Pi(\Phi) = -P(G\Phi),
\ee
with $(\partial \phi)_{i}\equiv \partial_{i} \phi_{i}$, $(G\Phi)_{i}\equiv G_{i}\Phi_{i}$ and $P(\Phi)$ the smeared version of $P_{a}^{i}$.
The algebra in terms of the $P,M$ variables is simply 
\be
{[}P(\Phi),M_{i}(x){]}_{q}= M_{i} \Phi_{i}(x), \quad {[}P(\Phi),P(\Psi){]}_{q} = P([\Phi,\Psi]).
\ee
Where the bracket index $q$ denotes quantum commutators to be distinguished from 
the Lie algebra commutators.
It will be useful to know also the commutator with the currents
\be\label{PJ}
{[}P(\Phi), J_{ij}{]}_{q}= H_{ij} \partial_{j}\phi_{j} H_{ij}^{-1} -D_{ij}\phi_{i}.
\ee
The generators of gauge symmetries can be naturally expressed in terms of these, if $X(x)=X^{a}(x)T_{a}$ labels
the parameter of the infinitesimal gauge transformation one can write
\be 
G_{X}=-\int \tr\left( X \nabla_{i}\frac{\delta}{\delta A_{i}}\right)= P(M_{i}XM_{i}^{-1}).
\ee
The generator of the holomorphic symmetry can be written in a similar form in terms of 
Lie algebra elements $H_{i}=H_{i}^{a}T_{a}$ satisfying $\partial_{i}H_{i}=0$,
\be
G_{H}= P(H_{i}).
\ee
One sees that the conditions of holomorphic invariant $G_{H}\sim 0$ is necessary
in order for the momentum $\Pi$ to be well defined since if $H_{i}$ is an holomorphic transformation
$\Pi(\partial H)= 0= P(H)$. 
Moreover the observables  $\Pi, J$ are gauge invariant 
\be
[G_{X}, P(\Phi)]_{q}
=[G_{X}, J_{ij}]_{q} =0.
\ee
When written in terms  these  observables the algebra reads 
\bes
{[}\Pi(\Phi),\Pi(\Psi){]}_{q}&=& -\Pi(\partial[G\Phi, G\Psi]),\\
{[}\Pi(\Phi), J_{ij}(x){] }_{q}&=& -(H_{ij}\Phi_{j}H_{ij}^{-1})(x) + (D_{ij}G_{i} \Phi_{i})(x),\label{pij}\\
{[}J_{ij}, J_{kl}{]}_{q}&=& 0.
\ees
where $[\Phi,\Psi]_{i}\equiv [\Phi_{i},\Psi_{i}]$ and
obviously $\partial[G\Phi, G\Psi]= [\Phi, G\Psi]+[G\Phi, \Psi]$.
 It might be interesting to note that this algebra possesses strong similarity with a centrally extended 
algebra, this is clear if one takes the derivative of (\ref{pij}) leading to
\be
{[}\Pi(\Phi), \partial_{i}J_{ij}{] }= \,[( G\Phi)_{i}, \partial_{i}J_{ij}] +  D_{ij}\Phi_{i}  -\partial_{i}(H_{ij}\Phi_{j}H_{ij}^{-1}) \label{pij2}.
\ee
%

\section{Determinant and regularization}\label{det}
 
The wave functions in the Schr\"odinger representation of pure Yang-Mills are gauge invariant functionals of $A_{i}$ and the scalar product is given by
\be
||\Psi||^2 = \int_{\cal{A}/\cal{G}} D\mu(A) \bar{\Psi}(A) \Psi(A),
\ee
where the integral is over the space of gauge connections modulo gauge transformations and 
$D\mu(A)= \frac{DA}{\mathrm{Vol(\cal{G})}}= \frac{\prod_{i}DA_{i}}{\mathrm{Vol(\cal{G})}}$.
From the previous section we know that we can equivalently describe gauge invariant wave functionals as holomorphic 
invariant wave functionals of $H_{ij}$ or $J_{ij}$.
In order to express the physical scalar product in this new variables we need to compute the jacobian.
Since $A_{i}= -\partial_{i}M_{i}M_{i}^{-1}$, 
$\delta A_{i}= -(\nabla_{i}\delta M_{i}) M_{i}^{-1} $, the change of variables involve a determinant
\be 
e^\Gamma\equiv \mathrm{det}\left(\frac{\delta A_{i}}{\delta M_{i}M_{i}^{-1}}\right)= \mathrm{det}(\nabla_{1}\nabla_{2}\nabla_{3}).
\ee
The variational derivative of the action is given by 
\be\label{delG}
\frac{\delta \Gamma}{\delta A_{i}^{a}(x)} =\tr_{ad}\left[(\nabla_{i})^{-1}(x,x)T_{a}\right]
\ee
the trace being in the adjoint representation.
Since $ \nabla_{i}= M_{i}\partial_{i}M_{i}^{-1}$ the covariant propagator can be expressed in terms of the standard propagator as a coincident limit
\be\label{detd}
\frac{\delta \Gamma}{\delta A_{i}^{a}(x)} = \lim_{x\to y} \left(M_{i}(x)G_{i}(x,y)M_{i}^{-1}(y)\right)^{bc}f_{abc}.
\ee
 A similar coefficient arises  when one expresses the kinetic term in terms of the group variables using (\ref{MA2})
\be\label{Del}
 \frac{\delta }{\delta A_{i}^{a}(x)}\frac{\delta}{\delta A_{i}^{a}(x)}
=   \lim_{x\to y} \left(M_{i}(x)^{-1}G_{i}(x,y)M_{i}(y) T_{b}\right)^{bc} (G_{i}P_{c}^{i})(y) + 
(G_{i}P_{a}^{i})(x) (G_{i}P_{a}^{i})(x).
\ee
Integration over $x$ and summations over indices should be understood.
The relation between these coefficients is not a coincidence,
indeed the presence of $e^\Gamma$ in the integration measure  insures that (\ref{Del}) is a self adjoint operator.

Of course, in order to make sense of these statements, one  needs a regularization of the propagator which preserves all the symmetries, namely the holomorphic symmetry
$M_{i}\mapsto M_{i} h_{i}^{-1}$. Since $\nabla_{i} = M_{i}\partial_{i}M_{i}^{-1} $ is invariant under holomorphic transformations this means that the regularized propagator
should   transform as $G_{i}^\mu(x,y)\mapsto h_{i}(x)G_{i}^\mu(x,y)h_{i}^{-1}(y)$, where $\mu$ denotes the momentum cut-off scale (this is obviously true for the 
unregulated propagator which satisfies $h_{i}(x)G_{i}(x,y)h_{i}^{-1}(y)=G_{i}(x,y)$).

The key ingredient in the construction of the regularized propagator are  the group valued functionals $\Lambda_{i}(x,y)$
defined to be
\be
\Lambda_{1}(x,y)\equiv H_{12}(y_{1},x_{2},x_{3})H_{23}(y_{1},y_{2},x_{3})H_{31}(y_{1},y_{2},y_{3}),
\ee
and  cyclic permutation for $\Lambda_{2},\Lambda_{3}$.
These group elements transform  under holomorphic transformation $H_{ij}\mapsto h_{i}H_{ij}h_{j}^{-1}$ as 
\be
\Lambda_{1}(x,y)\mapsto h_{1}(x_{2},x_{3})\Lambda_{1}(x,y)h_{1}^{-1}(y_{2},y_{3})
\ee
and similarly for $\Lambda_{2},\Lambda_{3}$.
They also satisfy the key properties 
\be\label{lprop}
\partial_{i}^x \Lambda_{i}(x,y) =0, \quad \quad  \mathrm{and}\quad \quad  \Lambda_{i}(x,x) =1.
\ee
We can now propose a regularization scheme which preserves  holomorphic invariance :
\be\label{defG}
{\bf G}_{i}^\mu(x,y) \equiv \int\extd z \,G_{i}(x,z) \Lambda_{i}(z,y)\delta_{\mu}(z,y)
\ee
where the regularised delta function is given by (say)
\be
\delta_{\mu}(x,y)= \prod_{i=1}^3\delta_{\mu}(x_{i}-y_{i}), \quad \delta_{\mu}(x_{i}) = \frac{\mu}{\sqrt{\pi}}e^{- \mu^2 x_{i}^2}.
\ee
This regularised propagator satisfies
\be
\partial_{i}^x {\bf G}_{i}^\mu(x,y) = \Lambda_{i}(x,y)\delta_{\mu}(x,y).
\ee
Moreover the integral (\ref{defG}) can be explicitly performed and one  gets
\be
{\bf G}_{1}^\mu(x,y)= \Lambda_{1}(x,y) \theta_{\mu} (x_{1}-y_{1})\delta_{\mu}(x_{2}-y_{2})\delta_{\mu}(x_{3}-y_{3})
\equiv  \Lambda_{1}(x,y) { G}_{1}^\mu(x,y),
\ee
where $\theta_{\mu}(x)= \frac1{\sqrt{\pi}}\int_{-\infty}^{\mu x}\extd t e^{-t^2}$ is one half plus half the error function.

Now that we have a regularization which preserves gauge invariance and holomorphic symmetry we can evaluate the propagator at coincident points 
and compute  ${\bf{G}}_{1}^\mu(x,x)= \frac{\mu^2}{2\pi}$. This propagator is proportional to the identity hence $\frac{\delta \Gamma}{\delta A_{i}}=0$
and 
\be 
|\det(\nabla_{1}\nabla_{2}\nabla_{3})| = |\mathrm{det}(\partial_{1}\partial_{2}\partial_{3})|
\ee
the determinant is independent of the connection. This  is independent on the form of the regulated delta function.
We have proven this results here in dimension $3$ but the  formalism is valid in any dimension therefore  in  
any dimension  $D$
\be 
|\det(\nabla_{1}\cdots\nabla_{D})| = |\mathrm{det}(\partial_{1}\cdots \partial_{D})|.
\ee

Since this is for us an important conclusion, let us give an other totally independent proof of the same result using lattice gauge theory.
Putting gauge theory on a lattice provides a gauge invariant regularization of the theory. 
We start with a gauge theory on a square periodic D-dimensional lattice, we choose one origin and an orientation and denote the displacement vector 
of one lattice unit in the direction $i$ by $e_{i}$, the sites of the lattice are labeled by $x= \sum_{i} x_{i}e_{i}$ with $x+N e_{i}$ being identified with $x$.
 The gauge connection is  encoded in terms of group elements $g_{i}(x)$ associated with the link $(x, x+e_{i})$.
The analogous of the Bars variables are given by holonomies starting at one corner of the lattice.
Let $x=\sum_{i} x_{i}e_{i}$, $0\leq x_{i}< N$, then
\be
M_{i}(x) =  g_{i}(x)^{-1}g_{i}^{-1}(x-e_{i})\cdots g_{i}^{-1}(x-x_{i}e_{i}),
\ee 
so that the starting element is always on the plane $x_{i}=0$.
We introduce the discrete derivative $\partial_{i} f (x) \equiv f(x)-f(x-e_{i})$, the $M_{i}$ satisfy the difference equation
\be
\partial_{i}M_{i}(x)M_{i}^{-1}(x)= (1- g_{i}(x)), \quad M_{i}(x)= 1, \, \mathrm{for}\, x_{i}=0.
\ee 
which is clearly the discrete analog of (\ref{defm}).
It can be equivalently written as $\nabla_{i}M_{i}=0$, where the discrete covariant derivative is 
$\nabla_{i}M(x)\equiv g_{i}(x)M(x)-M(x-e_{i})$.
Now, we want to compute the Jacobian of the transformation $g_{i}\to M_{i}$,  since 
$g_i({x})= M_{i}(x-e_{i})M_{i}^{-1}(x)$  we first remark that 
$M_{i}$ depends only on $g_{i}$ so that the transformation matrices is block diagonal and the problem is essentially 
one dimensional. A direct computation  gives the variation 
\be 
g_{i}(x)^{-1}\delta g_{i}(x) = (\nabla_{i}\delta M_{i}(x) )M_{i}^{-1}(x)= 
M_{i}(x) (\partial_{i}M_{i}^{-1}\delta M_{i}(x)) M_{i}(x).
\ee
Therefore 
\bes\nonumber
|\det(\nabla_{1}\cdots\nabla_{D})|&=& \prod_{i=1}^D |\det(\nabla_{i})|
=  \prod_{i=1}^D |\det(M_{i}\partial_{i}M_{i}^{-1})| \\
& =&  \prod_{i=1}^D |\det(\partial_{i})|=|\mathrm{det}(\partial_{1}\cdots \partial_{D})|
\ees
Where we have used the block diagonal form of the transformation matrix,  and also the factorization and invariance under 
conjugation of the determinants which are all valid operations since we are in a finite dimensional context.

There is an important subtlety here which is hidden in the domain of definition of the operator $\partial_{i}$.
This comes from the fact that $M_{i}(x)$ is not a periodic function on the lattice, since 
on one side of the lattice $M_{i}(0)=1$ and on the other side $M_{i}(Ne_{i})$ is arbitrary. This means that the determinant of 
$\partial_{i}$ which arises from the change of variables should be computed in the space of Lie algebra valued 
functions satisfying $\phi(0)=0$ and $\phi(Ne_{i})$ arbitrary and not on the space of periodic Lie algebra valued functions.
 One can easily see that then the residual derivative determinant can be exactly evaluated and is in fact equal to unity. 
This  simple answer can be traced back to the lattice regularization which starts from group variables and doesn't affect 
our main statement which is that the full determinant is independent of the connection.

\subsection{Gauge invariant measure}
This results allows us, up to a gauge field  independent determinant,
to make the change of variables from $A_{i}$  to $M_{i}$. The
induced measure ${\cal{D}}M_{i}$ on the fields $M_{i}$ should be thought 
as being the product over spacetime points of the Haar measure 
$ DM =\prod_x  dM(x)$.

More precisely, the proper way to define the measure is to give its moments evaluated on a proper class 
of functionals. The functionals which have a well defined integration under this  
 measure are called cylindrical functions $ F(M(x)) =F_n(M(x_1), ...,M 
 (x_n)) $ which depends on the value of the field only in a finite  
 number of points and the measure $DM$ is defined by its value on  
 cylindrical functions to be
 $\int DM F(M(x)) = \int dM_1...dM_n F_n(M_1,...,M_n)  $ the integral  
 being over a product over Haar measure.
The Haar measure is right-left invariant
and  one can write (since the measure is ultralocal, in the sense  just  defined, we can look at only a fix $x$)
$$\int dM_1 dM_2 dM_3 f(M_1, M_2, M_3) =
 \int dM_1 dH_{12} dH_{13} f(M_1,M_1 H_{12},M_1 H_{13})$$
  (the order of integration being important). If $f$ is gauge  
 invariant  the integral over $M_1$ factorises, supposing the measure on the unitary group  to be
 normalized we get
$ \int dH_{12} dH_{13} f(H_{12},H_{13}).$
 This shows that the measure on $A/G$ is  $$\int DH_{12} DH_{13} = \int DH_{{12}} DH_{13} DH_{23} \delta_{1}(H_{12}H_{13}H_{23})$$
where in the last equality we insert a delta function on the group to emphasize the symmetric form of the measure under  permutation
of indices. The interpretation of this delta function constraint should be clear: it is the integrated version of the Bianchi identity $\epsilon^{ijk}\nabla_{i}F_{jk}=0$
expressed in terms of gauge invariant observables.

 Now, the identity functional $\psi(H)=1$ can be viewed as a limit of identity  
 cylindrical functional the integral of which being always one if one choose the normalized Haar measure.
So we can conclude from this analysis that  the trivial wave functional is integrable.

This result is purely kinematical but it captures an essential feature of the formalism, namely the fact that 
even if we are working in the continuum, the choice of Bars variables allows us to recover results which are easily obtained in the lattice 
formulation, and as we will see, allows us to devise a strong coupling expansion.
If one computes the expectation  value of Wilson lines observables in this trivial kinematical vacuum one sees, for rectangular 
Wilson loops (\ref{Willoo}) and from the definition of the measure, that
\be \label{WR}
\langle W_{R}(y)\rangle = \int DH_{12} DH_{13} W_{R}(y) = \delta_{R,0}.
\ee
 Wilson loop expectation values are non zero only for trivial representation, this confirms the interpretation that 
this trivial vacuum corresponds to an infinite string tension, which is what one expects in a crude  $g\to \infty$ limit.
Now, one could say that such a result  seems at odds with the well known fact 
that in 2D  there is a non trivial determinant in the KKN formalism, namely the hermitian WZW action \cite{KN, KKN}.
This apparent contradiction can be easily resolved 
since, thanks to the work of Gawedzki and Kupianen \cite{GK}, we know that there is a deep 
relationship between hermitian and unitary model.   

More precisely if we specify their results to the genus zero case,
the  relation between correlation functions of WZW theory level $k$ 
 with $SU(N)$ group and field  $g(x)$ and the hyperbolic model with field 
$h(x)=m(x)^\dagger m(x)$ is given by
$$\int\prod\limits_{i=1}^N (g_{R_i})^{\alpha_i}_{\beta_i}(x_i)
\,e^{-kS(g)}\,Dg\ \int\prod\limits_{n=1}^N(h_{R_i})^{\beta_i}_{\alpha_i}(x_i)
\,e^{(k+2N)S(h)}\,\delta_1(h(x_0))\,dh\ =\ \cal{N}_V$$
where $\cal{N}_V$ is the number of conformal blocks (independent of $x_i$ and the sphere metric).
When extrapolated to $k=0$ clearly this identity shows the normalizability of the trivial wave functional.

Now it is also well known that the correlation functions  of the unitary WZW model are zero unless the 
highest weight $\Lambda_{R}$ is integrable, that is $(\Lambda_{R},\theta)\leq k $ where $\theta$ is the 
highest weight of the adjoint representation $(\theta,\theta)=2$.
If we extrapolate these results to $k=0$ this means that only primary fields associated to the trivial representation and its descendant are non zero
in agreement with (\ref{WR}). Moreover as noticed by KKN \cite{KKN} the correlators of the currents  of the unitary WZW model 
can be obtained by analytic continuation $(k+N)\to -(k+N)$ from the hermitian ones. The reason being that both
satisfy the KZ equations with opposite  $(k+N)$ and both are invariant combinations under monodromies,
which fix them uniquely up to normalization. 

\subsection{Semi-complex coordinates}
The determinant is trivial in the real basis where $H$ are unitary matrices. However, if one choose to work in the 
semi complex basis we  find a non trivial determinant which is a generalization of WZW model in higher dimension and 
is  related to the magnetic mass term introduced in \cite{NAlex}. 
For completeness we present this computation here.

We have seen in the first section that in the semi-complex coordinates $X=(x,\bar{x},u_{x})$ the gauge invariant data is entirely encoded into a complex 
$SL(N,\mathbb{C})$ field $H\equiv H_{uz}$.
The unregulated propagators are given by 
\bes
G_{u}(X,Y)&=& \theta(u_{x}-u_{y}) \delta^2(x-y),\\
G_{z}(X,Y)&=& \frac{1}{\pi(\bar{x}-\bar{z})}\delta(u_{x}-u_{y}),\\
G_{\bar{z}}(X,Y)&=& \frac{1}{\pi({x}-{z})}\delta(u_{x}-u_{y}).
\ees
The integrals (\ref{defG}) defining the regulated propagators can be explicitly performed leading to 
\bes
{\bf{G}}_{u}^{\mu}(X,Y)&=& \theta_{\mu}(u_{x}-u_{y}) \delta_{\mu}^2(x-y)\Lambda_{u}(X,Y),\\
{\bf{G}}_{z}(X,Y)&=&  \frac{\delta_{\mu}(u_{x}-u_{y})}{\pi(\bar{x}-\bar{y})}\left(1-e^{-\mu^{2}|x-y|^{2}}\Lambda_{z}(X,Y)\right),\\
{\bf{G}}_{\bar{z}}(X,Y)&=&  \frac{\delta_{\mu}(u_{x}-u_{y})}{\pi({x}-{y})}\left(1-e^{-\mu^{2}|x-y|^{2}}\Lambda_{\bar{z}}(X,Y)\right),
\ees
with $|x|^2=x\bar{x}$ and
\bes
\Lambda_{u}(X,Y) &=&H(x,\bar{x},u_{y})(H^\dagger H)^{-1}(y,\bar{x},u_{y})H^{\dagger}(y,\bar{y},u_{y}),\\
\Lambda_{z}(X,Y) &=&(H^\dagger H)^{-1}(y,\bar{x},u_{x})H^{\dagger}(y,\bar{y},u_{x})H(y,\bar{y},u_{y}),\\
\Lambda_{\bar{z}}(X,Y) &=&H^{\dagger}(x,\bar{y},u_{x})H(x,\bar{y},u_{y})(H^\dagger H)^{-1}(y,\bar{y},u_{y})\nonumber.
\ees
The value of the propagator at coincident point is given by
\bes
&{\bf{G}}_{u}^{\mu}(X,X)= \frac{\mu^{2}}{2\pi},& \\
&{\bf{G}_{z}}(X,X)= -\frac{\mu}{\pi^{\frac32}}(H^{\dagger}H)^{-1}\bar{\partial}(H^{\dagger}H),\nonumber
\quad {\bf{G}}_{\bar{z}}(X,X)= \frac{\mu}{\pi^{\frac32}}(\bar{\partial}(H^{\dagger}H)) (H^{\dagger}H)^{-1}.&
\ees
The variational derivative of the determinant (\ref{detd}) can be explicitly computed and after some algebra one gets
\be
\delta\Gamma = -\frac{2 N \mu}{\pi^{\frac32}} \int \tr\left\{(H^{\dagger}H)^{-1}\delta(H^{\dagger}H)\partial\left((H^{\dagger}H)^{-1}\bar{\partial}H^{\dagger}H\right)\right\}
\ee
This equation can be easily integrated out in terms of a three dimensional generalization of a Wess-Zumino-Witten action
\be\nonumber
-\frac{\pi^{\frac32}}{2N\mu}\Gamma= \frac1{2}\int \extd u \extd^{2} z\tr\left(\partial(H^{\dagger}H) \bar{\partial}(H^{\dagger}H)^{-{1}}\right) +
 \frac{i}{12}\int \extd u \int_{B_{u}} \tr\left([(H^{\dagger}H)^{-1}d(H^{\dagger}H)]^{3}\right).
\ee
The last integral being over a three ball $B_{u}$ bounding the plane $u=cste$.

\section{Hamiltonian}\label{ham}

In this section we write down the Hamiltonian in terms of gauge invariant variables.
From now on and in the following one shall stick to three (space) dimensions, 
a real basis and we 
introduce some notations well adapted to this case.
The vacuum wave functional we look for, is expressed as a functional of the currents $J_{12}, J_{23}, J_{31}$.
These should be thought as a vector (or three dimensional holomorphic connection) and we denote
\be
H_{i} \equiv H_{i-1 i}, \quad J_{i}\equiv J_{i-1 i}=
\partial_{i}H_{i} H_{i}^{-1},\quad D_{i} \equiv D_{i-1 i}=\partial_{i}-J_{i}
\ee
and we introduce the `magnetic field'\footnote{It is related to the true magnetic field $F_{i}$ by
$B_{i-1}= -M_{i}^{-1} F_{i-1} M_{i}$}
\be
B_{i-1}= \partial_{i}J_{i+1}.
\ee
(all indices are modulo three indeed)
The potential term is readily expressed in these variables 
\bes
\mathcal{V} &=& -\sum_{i}\int \tr\left((\partial_{i}J_{i+1})^2(x)\right) \extd x = -\sum_{i}\int \tr (B_{i}(x))^2 \extd x\\
&=&\frac12\sum_{i,a}\int B_{i}^{a}(x)B_{i}^{a}(x) \extd x
\ees
and with the help of the holomorphic regularization  one can compute the kinetic term (\ref{Del}): the coefficient of the linear term  
vanishes for reasons identical to the vanishing of the variation of the jacobian and we are left with
\be\label{T}
\mathcal{T} = -\frac12\sum_{i}\int (G_{i}^tP_{a}^{i})^{\dagger}(x) (G_{i}^tP_{a}^{i})(x) \extd x
= \frac12 \sum_{i}\int P_{a}^{i}(y) \tilde{\Theta}^{ab}_{i}(y,z) P_{b}^{i}(z) \extd y \extd z
\ee
Here for later convenience we have introduced a (unregularized at this stage) kernel
\be
\tilde{\Theta}_{i}^{ab}(y,z) = \delta^{ab}(G_{i}G_{i}^t)(y,z)= \delta^{ab}\int G_{i}(y,x)G_{i}(z,x) \extd x.
\ee
and we have used that $(G_{i}^tP_{a}^{i})^{\dagger}(x)= (P_{a}^{i} G_{i})(x)$.

If one look at the detailed structure of the kinetic term
one encounters a disturbing  infrared divergence hidden in it, namely if one look at (\ref{T}) the structure of each term is given by
\be
-\frac12 \int \extd^3 x\extd y_{1}\extd z_{1} \theta(y_{1},x_{1}) \theta(z_{1},x_{1}) P_{1}^{a}(y_{1},x_{2}, x_{3}) P_{1}^{a}(z_{1}, x_{2},x_{3})
\ee
using the identity $1=\theta(x,y)+\theta(y,x)$ we can express this term as 
\bes
& & -\frac12 \int\extd x_{2}\extd x_{3}\left(\int_{x_{1}}^{+\infty} P_{1}^{a} \extd y_{1}\right) \left(\int_{-\infty}^{+\infty}P_{1}^{a} \extd z_{1}\right)(x_{2}, x_{3})\label{1}\\
& & + \frac12 \int \extd^3 x\extd y_{1}\extd z_{1} \theta(y_{1},x_{1})\theta(x_{1},z_{1})P_{1}^{a}(y_{1},x_{2}, x_{3}) P_{1}^{a}(z_{1}, x_{2},x_{3}).\label{2}
\ees
The first term in this expansion contains $$\int_{-\infty}^{+\infty}P_{1}^{a}(y_{1},x_{2}, x_{3}) \extd y_{1} = \int \tr(P_{1}(y) \phi^{a}_{x}(y)) \extd y$$
which is a generator of an infinitesimal holomorphic symmetry  with generator $\phi^{a}_{x}(y) = -2T^{a}\delta(x_{2}-y_{2})\delta(x_{3}-y_{3})$, $\partial_{1}\phi^{a}_{x}(y) =0$
and therefore acts trivially on holomorphic invariant states. We can therefore drop this term and consider the kinetic term 
to be given by the second term (\ref{2}). This amounts to replace in all the previous expressions $\tilde{\Theta}_{i} = GG^t$ by the convolution product
$\Theta_{i} =GG$. This second term is now free of any infrared problem, 
the integral over $x_{1}$ can be performed and we are left with 
\be
-\frac12\int\extd y_{1}\extd z_{1} \extd x_{2} \extd x_{3} |y_{1}-z_{1}|  \tr(P_{1}(y_{1},x_{2}, x_{3}) P_{1}(z_{1}, x_{2},x_{3}) )
\ee
which exhibits a stringy nature of the kinetic term with some linearly rising potential.
There is even a deeper justification arguing also in favor of  this second form of the kinetic term which comes from the derivation, in the connection variables, of its 
 matrix elements  given by 
\be
\langle\Psi |\mathcal{T}|\Psi\rangle 
=\int_{\cal{A}/\cal{G}} DA \overline{\left(\frac{\delta {\Psi}(A)}{\delta A_{i}^{a}}\right)}\frac{\delta {\Psi}(A)}{\delta A_{i}^{a}}.
\ee
We also want our theory to be CPT invariant which means that instead of conjugating with simple complex conjugation $\Psi \to \overline{\Psi}$
we can equivalently replace this conjugation by a CPT transformation $\Psi \to CPT({\Psi})$.
The CPT transformation of the gauge field is given 
 $CPT(A_{i}(x))= -A_{i}^{\dagger}(-x)= A_{i}(-x)$.
When we change variables from $A_{i}$ to $M_{i}$ we need to integrate over one dimension from $-\infty $ to $x$, this breaks parity 
and the CPT conjugate of $M_{i}$ is not just the hermitian conjugate : it is given by a path ordered integral which starts from $+\infty$
that is $CPT(M_{i}(x)) = (M_{i}(-x) h_{i})^{\dagger}.$ Where 
$h_{1}(x_{2},x_{3})= \overleftarrow{e}^{\int_{-\infty}^{+\infty}A_{1}(-x_{1},-x_{2},-x_{3}) dx_{1}}$, is the holonomy across space at fixed 
$x_{2},x_{3}$, and $\partial_{i}h_{i}=0$.
 Of course if the  wave function is holomorphic invariant it doesn't depend on $h_{i}$ and CPT invariance is satisfied.
Now if one starts from a form of the scalar product with is manifestly CPT invariant (that is $CPT(\Psi)$ is used instead of $\overline{\Psi})$ and compute from there 
the form of the kinetic term one can see that the $A$ derivative acting on the right can be expressed in terms of the $M$ and the propagator $G$
but the derivatives acting on the left should be expressed in terms of the propagator $G^{t}$ and in this case 
the form of the kinetic term is expressed in terms of $\Theta= GG$. This form of the kinetic term being the one in which CPT invariance is manifest.
That is the kinetic term we will use is 
\be\label{T2}
\mathcal{T}
= \frac12\sum_{i,a,b}\int P_{a}^{i}(y) {\Theta}_{i}^{ab}(y,z) P_{b}^{i}(z) \extd y \extd z
\ee
Here the unregularised kernel is 
\be
{\Theta}_{i}^{ab}(y,z) = \delta^{ab}(G_{i}G_{i})(y,z)= \delta^{ab}\int G_{i}(y,x)G_{i}(x,z) \extd x.
\ee
This kernel satisfies 
\bes
&\partial^{y}_{{i}}\Theta_{i}^{ab}(y,z)=  \delta^{ab}G_{i}(y,z),
\qquad \partial^{z}_{i}\Theta_{i}^{ab}(y,z)= - \delta^{ab}G_{i}(y,z),&\\
&\partial^{y}_{{i}} \partial^{z}_{{i}}\Theta_{i}^{ab}(y,z)= -\delta^{ab}\delta(y,z).&
\ees
We can now express the kinetic term in the current variables, first recalling for comfort and according to our new notations 
the commutator   
\be
[P_{a}^k(x), J_{i}^{b}(y)]= \delta_{i}^k (H_{k}(y))^{b}{}_{a}\partial_{k}^{y}\delta(x,y)
- \delta_{i}^{k+1}(D_{k+1}^y)^{b}{}_{a} \delta(x,y).
\ee
Therefore 
\be\label{T3}
\mathcal{T} = \frac12\sum_{i,j,b,c}\int \extd y \extd z\,  \Omega_{ij}^{bc}(y,z) \frac{\delta}{\delta J_{i}^b(y)} 
\frac{\delta}{\delta J_{j}^c(z)}
\ee
with
\bes\label{om}
\Omega_{ii}^{bc}(y,z) &=& \left[(D_{i}^y)^{ba}(D_{i}^z)^{ca} \Theta_{i-1}(y,z) - \delta^{bc}\delta(y,z) \right],\\
\Omega_{i+1i}^{bc}(y,z) &=&\frac12 \left[D^{y}_{i+1}H_{i}^{-1}(z) \right]^{bc}(G_{i}(y,z)-G_{i}(z,y)).
\ees
The full QCD hamiltonian is
\be\label{H}
\mathcal{H} = g^2 \mathcal{T} + \frac{1}{g^2}\mathcal{V}.
\ee

\subsection{Regularised Hamiltonian}
In order to define mathematically the theory we need to include a regulator. 
Preserving holomorphic invariance is a strong constraint on the form  of the regulator 
and  we have seen that there is a natural holomorphic invariant regularization with scale parameter $\mu$ which is available in our context.
This regularization amounts to replace $\delta^{ab}G_{i}(x,y)$ by its holomorphic invariant regulated version 
(\ref{defG}) ${{\bf G}_{i}^\mu}^{ab}(x,y)$, the 
holomorphic regularization of the kernel  $\Theta$
is denoted $\Theta^\mu$ and is given by 
\be
\Theta_{i}^{\mu bc}(y,z)\equiv ({\bf G}_{i}^{\mu }{\bf G}_{i}^{\mu})^{bc}(y,z)=\int \extd x 
\left(\Lambda_{i}(y,x)\Lambda_{i}(x,z)\right)^{bc} G_{i}^{\mu}(y,x)G_{i}^\mu(x,z)
\ee
with 
$$
G_{1}^\mu(y,x)=  \theta_{\mu} (y_{1}-x_{1})\delta_{\mu}(y_{2}-x_{2})\delta_{\mu}(y_{3}-x_{3}),
$$
and  (see section \ref{det}),
\be\label{Ldef}
\Lambda_{1}(y,x)= H_{2}(x_{1},y_{2},y_{3})H_{3}(x_{1},x_{2},y_{3})H_{1}(x_{1},x_{2},x_{3})
\ee
  plus cyclic permutations.  Under holomorphic transformations we have 
\bes\nonumber
& M_{i}\to M_{i} h_{i}^{-1}, \quad H_{i}\to h_{i-1}H_{i}h_{i}^{-1}, 
\quad P^{i}_{a}\to (h_{i})_{a}{}^{b}P^i_{b}, \quad B_{i-1}\to h_{i}B_{i-1}h_{i}^{-1} &\\
&\Lambda_{i}(y,x)\to h_{i}(y)\Lambda_{i}(y,x)h_{i}(x)^{-1} &
\ees
and 
$\Theta_{i}^\mu(y,z)\to h_{i}(y)\Theta_{i}^\mu (y,z)h_{i}(z)^{-1}$.

The regularised kinetic term written in a manifestly hermitian form is given by
\bes\label{Treg}
\mathcal{T}
&=&  \frac14\sum_{i,a}\int ({\bf G}_{i}^{\mu t}P^{i})_{a}^{\dagger}(x) ({\bf G}_{i}^{\mu }P^{i})_{a}(x)+  ({\bf G}_{i}^{\mu }P^{i})_{a}^{\dagger}(x) ({\bf G}_{i}^{\mu t}P^{i})_{a}(x)  \extd x, \\
&=& \frac14\sum_{i,b,c}\int P_{b}^{i}(y) ({\Theta}_{i}^{\mu bc}(y,z) +{\Theta}_{i}^{\mu cb}(z,y))P_{c}^{i}(z) \extd y \extd z, \\
&=& \frac14\sum_{i}\int \left[P_{}^{i} ({\Theta}_{i}^{\mu} +{\Theta}_{i}^{\mu t})P_{}^{i}\right](y,z)  \extd y \extd z.
\ees
where we used the transposed kernels $$({\bf G}_{i}^{\mu t})^{ab}(x,y)\equiv -({\bf G}_{i}^{\mu })^{ba}(y,x), \quad 
({\Theta}_{i}^{\mu t})^{ab}(x,y)\equiv ({\Theta}_{i}^{\mu })^{ba}(y,x).$$

It is convenient for our subsequent computations to move all the derivatives to the right; in doing so we  
encounter a commutator
\be
[P_{i}^{b}(y), \Theta^{\mu bc}(y,z)]\equiv \lim_{{\tilde \mu}\to \infty}[P^{i}_{{\tilde \mu}b}(y), \Theta^{\mu bc}(y,z)],
\ee
where $P^{i}_{{\tilde \mu}}$ is a left invariant derivative regulated  according to our prescription: it satisfies
  $({\bf G}_{i}^{{\tilde \mu}}P^{i})_{a} =(G_{i}P_{{\tilde \mu}}^{i})_{a}$ and is explicitly given by
\be
P^{i}_{{\tilde \mu} a}(y)\equiv \int \extd z \Lambda_{i}^{ab}(y,z) \delta_{{\tilde \mu}}(y,z) P^{i}_{b}(z).
\ee 
The calculation of this commutator  gives
\be
\int [P_{b}^{i}(y), \Theta^{\mu bc}_{i}(y,z)] \extd y = -\Theta^{\mu c}_{i}(z,z),
\ee
where $\Theta^{c}\equiv \Theta^{ab}f_{ab}{}^{c}$.
This term is of order $O(\frac1{\mu^{2}})$. For the other commutator we find 
\be
\int [P_{b}^{i}(y), \Theta^{\mu cb}_{i}(z,y)] \extd y = \rho_{i}^{c}(z),
\ee
with 
\be
\rho_{i}^{c}(z) =\int\extd x ({\bf G}_{i}^{{\mu}}(z,x){\bf G}_{i}^{{\mu}}(x,z_{x_{i}}))^{c}
\ee
where $z_{x_{1}}=(x_{1},z_{2},z_{3})$, $z_{x_{2}}=(z_{1},x_{2},z_{3})$ etc...
This term is also of order $O(\frac1{\mu^{2}})$ and independent of $z_{i}$, it corresponds to an holomorphic transformation.
%
%
%
%
 We can now write the kinetic term in terms of the currents : in order to do so one needs to evaluate the 
double commutator 
\be
\int  \Theta^{\mu bc}_{i}(y,z) [P_{b}^{i}(y),[P_{c}^{i}(z), J_{k}^{a}(x)]] \extd y \extd z
\ee
which is also of order $O\left(\frac1{\mu^2}\right)$.
The conclusion is that up to terms $\int \extd z \Omega_{i}^c(z) \frac{\delta}{\delta J_{i}^c(z)}$ 
with $\Omega_{i}^c(z) = O\left(\frac1{\mu^2}\right)$ the regularised kinetic term takes the form (\ref{T3}) with
\bes\label{omreg}
\Omega_{ii}^{bc}(y,z) &=&\left[{D}_{i}^y\bar{D}_{i}^z \Theta_{i-1}(y,z) + H_{i}(y)(\partial_{i}^y\partial_{i}^z \Theta_{i} (y,z))(H_{i}(z))^{-1}\right]^{bc},\\
\Omega_{i+1i}^{bc}(y,z) &=&-\frac12\left[D^{y}_{i+1}H_{i}^{-1}(z) \partial_{i}^z\left(\Theta_{i}(y,z)+\Theta_{i}^{t}(y,z)\right)\right]^{bc},
\ees
where $(\Theta^t)^{ab}(x,y)=\Theta^{ba}(z,x)$ and $({D}\bar{D} \Theta)^{bc}\equiv {D}^b{}_{a}{D}^c{}_{d} \Theta^{ad}$.
\section{Vacuum wave functional}\label{vac}

In this section we show how one can, in our formalism, determine the behavior of the vacuum wave function both in the weak and strong coupling regime.
\subsection{Weak coupling regime}

Let us first look at the weak coupling limit of the theory : as is usual one first rescale the 
currents $J\to J/ g$ and then expand the hamiltonian in power series in $g$, $\mathcal{H}= \sum_{i}g^{i}\mathcal{H}_{i}$.
After field rescaling, the potential term $\mathcal{V}$ is of order zero, therefore this 
 expansion can be realized entirely as an expansion in the kinetic term and more precisely in the coefficients 
$\Omega$.
This expansion is an expansion in the number of currents $J$. Since $J$ has a dimension of mass 
it always enters the expansion in the combination $J/p$, if one think in terms of Fourier modes. This means that we 
should interpret this expansion as  being valid in the ultraviolet sector $p\to \infty$ of the theory and $\mathcal{H}_{0}$
is the Hamiltonian governing the gluon dynamic in the deep ultraviolet, this Hamiltonian being free.

The diagonal terms $\Omega_{ii}$ contains only a finite number of terms up to $g^4$ but the off diagonal component 
due to the presence of  the group element $H$ contains an infinite number of terms in this expansion.
We are now interested only in the structure of the first non trivial term $\mathcal{H}_{0}$
and the coefficients of the kinetic term at zero order are  given by 
\bes
\Omega_{ii}^{(0)bc}(y,z) &=& \left[(\partial_{i}^y)(\partial_{i}^z) \Theta_{i-1}(y,z) - \delta(y,z) \right]\delta^{bc},\\
\Omega_{i+1i}^{(0)bc}(y,z) &=& \frac12\left[\partial^{y}_{i+1} {(G_{i}(y,z)-G_{i}(z,y)})\right]\delta^{bc}.
\ees
This expression is not particularly illuminating at first sight, however if one 
writes down this hamiltonian in Fourier space $B(p)=\int \extd x e^{-ipx} B(x)$ and takes as fundamental variables 
the magnetic fields $B_{i}=\partial_{i+1}J_{i-1}$, after simple algebra the hamiltonian  becomes 
\be\nonumber
\mathcal{H}_{0}= \frac12 \int \frac{\extd^3 p }{(2\pi)^3}\left[ -(2\pi)^6 \left(p^{2}\delta_{ij}-p_{i}p_{j}\right) \frac{\delta}{\delta B_{i}^a(p)} 
\frac{\delta}{\delta B_{j}^a(-p)} +    B_{i}^a(p) B_{i}^a(-p)\right].
\ee
The Poincar\'e invariance of the theory becomes manifest.
The matrix contracting the derivatives has eigenvalues $p^2, p^2, 0$
the eigenstates associated to $p^2$ correspond to the two gluons polarization  
perpendicular to $p$. The eigenstate corresponding to the eigenvalue $0$ is puzzling at first sight since we are working in a gauge 
invariant formalism. In order to properly understand it one needs to remember that the gauge invariant measure constructed previously
contains only two scalar independent variables per Lie algebra generator, this is encoded in the constraint $H_{1}H_{2}H_{3}=1$
or in differential form in the Bianchi identity $\nabla_{i}F^{i}=0$.
This means that $J_{1}, J_{2}, J_{3}$ are not independent variables, and that any operator proportional to this constraint is strongly $0$.
When expressed in terms of our local variables and in the limit $g\to 0$ this constraint imposed by the measure reads 
$\sum_{i} p_{i} B_{i}(p)=0$.

We can now easily diagonalize the free hamiltonian $\mathcal{H}_{0}$ subject to this constraint and 
give explicitly the ground state in this regime
\be
\Psi(B) = \exp -\frac1{2g^2}\int \frac{\extd^3 p }{(2\pi)^3} \frac{ B_{i}^a(p) B_{i}^a(-p)}{|p|}
\ee
which is indeed the one expected and where we have reinstalled the dependence in $g$ by rescaling to the original variables.

\subsection{Evidence for confinement}

The previous result  is not very impressive by itself  since it just shows that our formalism can handle 
appropriately the perturbative regime which is well understood by other means.
However, if one look back at the formulation in terms of Bars variables 
this formalism, even if it is local, is very similar to the lattice formulation of QCD which is easily amenable to strong coupling expansion.
It shows that the route we have chosen allows us to deal also with  the strong coupling or infrared regime of the theory.

In the strong coupling regime $g\to \infty$ the hamiltonian $H$ is dominated by its kinetic term, as emphasized for instance in the hamiltonian formulation of 
lattice QCD \cite{KS}.
At first order in the strong coupling expansion the vacuum state is just the constant wave functional which is normalizable as we have seen even if we work in the continuum.
In order to go beyond this naive result and construct the vacuum state in the infrared regime, 
we  establish as a first key result that the potential term $\mathcal{V}$ viewed as a wave functional is an eigenstate 
of the kinetic term.
Since the eigenvalue  has a dimension of a mass we see the emergence of a dynamically generated mass scale, namely
\be \label{result}
 g^2\mathcal{T} \cdot  :\mathcal{V}: =  M  :\mathcal{V}: \quad \mathrm{with }\quad  M =  \frac{g^2 N \mu}{2(2\pi)^{\frac32}},
\ee
where $N$ is the dual coxeter number $N\delta^{a}_{b} = f^{apq}f_{bpq}$ (with our convention on Lie algebra generators) 
the $: :$ denotes normal ordering of the operator and $\cdot$ denotes the action of the operator $\mathcal{T} $ 
and $\mu$ is our ultraviolet cutoff regulator. It is important to keep in mind that this result is valid only 
in the strong coupling regime, the true  QCD mass scale  should appear in the continuum limit as the regulator is removed $\mu\to \infty$.

The first step of the computation follows directly from the definition of the regulated Hamiltonian 
(\ref{T3}, \ref{omreg})
which is defined in terms of  the 
holomorphically invariant regulated $\Theta^{\mu}_{i}(y,z)^{ab}$ and its derivative.
From this definition, one obtained after integration by part and noticing that the off diagonal terms do not play any role
\bes
&& g^2\mathcal{T} \cdot \int -\tr(\partial_{i-1}J_{i})^2(x) \, \extd x \label{TV}\\
&&  = \frac{g^{2}}{2} \int \left[\partial_{i-1}^y \partial_{i-1}^z\left(D_{i}^y\bar{D}_{i}^{z} 
\Theta_{i-1}^\mu(y,z) + H_{i}(y)(\partial_{i}^y \partial_{i}^z \Theta_{i}^\mu(y,z)) (H_{i}(z))^{-1}\right)^{bb}\right]_{y=z}\extd y.\nonumber
\ees
The RHS of (\ref{TV}) is the sum of two terms which can be computed separately, this relatively lengthy calculation is presented in detail in the  appendix.
The computation of the first term  of  (\ref{TV}) gives 
\bes
&&-\left[
\partial_{i-1}^y \partial_{i-1}^z 
           \tr_{ad} \left(
                            D_{i}^y\bar{D}_{i}^{z} \Theta_{i-1}^\mu(y,z) 
                            \right)
\right]_{y=z}
= \\ \nonumber
&& \qquad \qquad \qquad\frac{\mu}{4(2\pi)^{\frac{3}{2}}} \,\left[ -2 N\tr(B_{i+1}^2) - 4\mu^2(N^2-1) \right] +  O\left(\frac{1}{\mu}\right).\label{id2}
\ees
The computation of the second term  of  (\ref{TV})  gives the same result !
\bes\label{id1}
&&-\left[\partial_{i-1}^y \partial_{i-1}^z\tr_{ad}\left(H_{i}(y)(\partial_{i}^y \partial_{i}^z\Theta_{i}^\mu(y,z))(H_{i}(z))^{-1}\right)\right]_{y=z} = \\
&&\qquad \qquad \qquad \frac{\mu}{4(2\pi)^{\frac{3}{2}}} \,\left[ -2 N\tr(B_{i+1}^2) - 4\mu^2(N^2-1) \right] + O\left(\frac{1}{\mu}\right).
\nonumber \ees
The factor $N$ in front of the trace is the dual coxeter number coming from the relation between trace in the adjoint and in the vectorial
$\tr_{ad}(T^{a}T^{b})= 2 N \tr(T^{a}T^{b})$.
Summing over $i$ shows that (\ref{TV}) is 
\be
\frac{g^2N\mu}{2(2\pi)^{\frac{3}{2}}}\left(-  \sum_{i}\tr(B_{i}^2) -6\mu^2\frac{N^2-1}{N} \right) +  O\left(\frac{1}{\mu}\right).
\ee
Therefore if one defines the normal ordered operator
\be: \mathcal{V} : \equiv \mathcal{V} - 6 \mu^2 \frac{N^2-1}{N},
\ee
then one gets the announced result (\ref{result}).
It is interesting to note that $\mathcal{T}$ acts diagonally not only on the Lorentz invariant combination $\mathcal{V}$ but also on each term in the sum that is 
\be
\mathcal{T} \cdot  :\tr(B_{i}^2): = M  :\tr(B_{i}^2):, \quad   :\tr(B_{i}^2):\equiv  \tr(B_{i}^2) +2 \mu^2 \frac{N^2-1}{N}.
\ee
This result is extremely simple and strongly resonates with the 2 dimensional results.
It suggests that the action of the kinetic term on local gauge invariant operators is analogous to a kind of dilation operator which 
counts the number of $J$ constituents of the operators it acts on. 
In order to validate such a picture higher order computation should be performed.
 
\subsection{strong coupling expansion}

We can now exploit the full strength of our formalism and show that
the strong coupling expansion can be performed following a strategy devised in 2+1 by KN and strongly reminiscent of the 
cluster expansion of Lattice QCD \cite{clusterexp}.
The QCD hamiltonian reads (lets take a finite dimensional analog for the presentation of the argument)
\be
H= g^2 T +\frac1{g^2} V , \mathrm{with}\,\, T= \Omega^{ij}\partial_{i}\partial_{j}
\ee 
Then we see that in the strong coupling regime $g\to \infty $
the hamiltonian reduces to its kinetic term and the trivial wave function $\Psi= 1$ is a solution 
in this regime to the equation $H\psi =0$ up to order $1/g^{2}$, this solution is normalisable as we have seen in the previous section.
In order to go further 
lets suppose that 
\be\label{diagT}
\Omega^{ij}\delta_{i}\delta_{j}V = M V,
\ee
which is what we have proven in the last section in the context of QCD, 
and lets 
consider the wave function $\Psi = e^{- \frac{1}{g^2M} V } \Phi$, using the previous equality we have 
\be
H  e^{- \frac{1}{g^2M} V } \Phi =e^{- \frac{1}{g^2M} V }   \tilde{H}\Phi,
\ee
where 
\be
\tilde{H} = g^2 T - \frac2M \Omega^{ij} \partial_{i}V \partial_{j} + \frac{1}{g^4 M^2} \Omega^{ij} \partial_{i}V \partial_{j}V
\ee
and we see that the wave function $\Phi =1$ or $ \Psi = e^{- \frac{1}{g^2M} V }$  is a vacuum solution at order 
$1/g^4$ and therefore describes the vacuum in the strong coupling regime.

The higher order terms in this expansion can be recursively constructed in principle. For instance at the next order 
the vacuum wave functional reads  $\Psi = e^{- \frac{1}{g^2M}( {\cal V} + \frac{P}{M^2})} $ where $P$ satisfy the equation
\be 
\frac{g^{2}}{M}[ [{\cal T}, {\cal V}], {\cal V}]1 \equiv\frac{g^{2}}{M}{\cal T}\cdot P = \frac12 \int \tr(F_{i}\Delta F_{i})-\epsilon^{ijk}\tr(F_{i}[F_{j},F_{k}])
\ee
with $\Delta$ the covariant laplacian.
Now one can see from the definition of the regularised kinetic term that   $\frac{g^{2}}{M}{\cal T}$
when acting on gauge invariant operators of order $6$  reproduces a linear combination of such operators 
plus a sum over lower dimensional terms $\mu^2{\cal V}$ and $\mu^{6}$ which contributes to a renormalization of the vacuum energy and coefficients of 
${\cal V}$. Thus up to renormalization, which can be reabsorbed at this order in a redefinition of $M\to M(1+\frac{\alpha}{g^4N^2})$, with $\alpha$ a numerical constant,
we have 
\be
P= \int a\tr(F_{i}\Delta F_{i})-b\epsilon^{ijk}\tr(F_{i}[F_{j},F_{k}]).
\ee
where $a,b$ are numerical coefficients (independent of $N,\mu,g$). These coefficients are uniquely computable from 
the regularization prescription we have given. This calculus is quite lengthy but otherwise straightforward and will be presented elsewhere.
This procedure can be implemented at higher order in a strong coupling expansion.

%
The conclusion is then that in the infrared regime  ( more precisely for slowly varying  and 
low amplitude field) and in a strong coupling expansion the vacuum wave functional is given by
\be\label{vaccua1}
\Psi = e^{- \frac{1}{g^2m}\int \tr(B^2)}.
\ee
where the mass scale $m$ computed here only to first order in $1/g^4N^2$ can be recursively computed in a strong coupling expansion.
 That is $m= M(1+\frac{\alpha}{(g^2N)^2} + \cdots)$ where the dots refer to terms of higher order 
in $1/(g^{2}N)^2$.
We can now give easily the physical interpretation of the dynamical scale $m$ entering the definition of the infrared 
vacua wave functional \cite{dimred}.
Lets compute the expectation value of a large Wilson loop $W_{C}$ with area $A_{C}$
$$\langle W_{C}\rangle= \langle\Psi_{vac}| W_{C}|\Psi_{vac}\rangle.$$
This average  should be  dominated  in the limit $A_{C}\to \infty$ by the expectation value with respect to the 
infrared vacua (\ref{vaccua1})  and is therefore given by 
\be
\langle W_{C}\rangle\sim \int_{A/G} DA e^{- \frac{2}{g^2 m}\int \tr(B^2)} W_{C}(A).
\ee
We recognize the partition function of 2+1 Euclidean Yang-Mills theory with coupling $g^2_{2+1}= \frac{g^2m}{2}$.
Now
 this expectation value reduces in the limit of large area to the expectation 
value of the Wilson line in the $2+1$ Yang-Mills vacua.
This vacua has been recently exactly constructed at least in the large $N$ limit where no screening due to $n$-ality is expected \cite{nality}
and since  $2+1$ Yang-mills shows the property of confinement 
one gets
\be
\langle W_{C}\rangle\sim  e^{-{\sigma} A_{C}}
\ee
where  the string tension is given by
\be
{\sigma}= g^4_{2+1}\frac{N^2-1}{8\pi}.
\ee
Translated back into $3+1$ Yang-mills coupling one gets 
\be
{\sigma}=  m^2 g^4\frac{N^2-1}{32\pi}.
\ee
One should insist here that this results has been only establish firmly only in a strong coupling regime 
at first order, that is $m=M$, and is not yet shown to survive the continuous limit.

\section{Conclusion}

We have seen in this work that is is possible to give a formulation of pure QCD in terms of local gauge invariant observables
and computed explicitly the jacobian for this transformation which  shows that the trivial wave function is normalisable.
We have also constructed the regularised hamiltonian and expressed the vacuum wave function at first order in a strong coupling expansion.
This expansion can be extended to higher order as an expansion in terms of higher dimensional local operators.

Several problems in this framework have not been explored yet. The most formidable is the problem of the continuum limit.
As we have seen a mass scale makes its appearance in the construction of the vacuum wave functional but this scale is cutoff dependent.
In order to establish on a robust basis the existence of a dynamical mass scale in the infrared one would need to show that it is possible to remove 
the cut-off without destroying the presence of a non zero string tension. 
This would mean first showing that no phase transition makes its appearance while going from strong to weak coupling 
(that is $M>0$ from strong to weak coupling) and moreover that the scaling 
of the dynamical mass is such that its continuum value is controlled by the usual renormalization group scaling \cite{UV}
\be
\frac{(4\pi)^2}{g^2N}= \frac{(4\pi)^2}{g^2(\Lambda)N} + \frac{11}{3}\ln\left(\frac{\mu^2}{\Lambda^2}\right)+
 \frac{34}{11}\ln\left(\ln\left(\frac{\mu^2}{\Lambda^2}\right)\right).
\ee
Notice that the detail study of the continuum limit in 2+1 pure QCD from the KKN approach 
needs to be completed too.

A related project which is less ambitious  but phenomenologically more interesting concerns the possibility to construct in 3+1 pure Yang-Mills 
the  kernel interpolating between the infrared and ultraviolet behavior of the vaccuum wave functional and extract from this 
an estimate for the mass gap in terms of the string tension \cite{jointpaper}. 

{\bf{Acknowledgments:}}
First, I would like to thank Djordje Minic and Rob Leigh for numerous discussion and a collaboration
which is the raison d'\^etre of this project.
We are very grateful to Lee Smolin for his strong support and we would like to thank K. Gawedzki for numerous exchanges on 
the physics of QCD and WZW.
We would like to thank A. Yelnikov, E. Livine, S. Speziale for their help and the physicists of Perimeter institute for numerous discussions.

\section{appendix}

In this appendix we want to establish the main technical result of this paper, that is the proof of the fact that 
the kinetic term acts diagonally on the potential term which is the lowest order local gauge invariant operator 
(\ref{result}). This amounts, has shown in the main body of the paper, to the computation of 
\be\label{main}
\partial_{i-1}^{y}\partial_{i-1}^{z} \tr_{Ad}\left(\Omega_{ii}^\mu(y,z)\right)|_{z=y}.
\ee
In order to perform this calculation it is convenient to introduce some conventions and establish some basic results.
First one  recall the definition of the current $J_{i}$
\be \label{dJ1}
\partial_{i}H_{i}= J_{i}H_{i},
\ee
One also introduce an auxiliary current $\tilde{J}_{i}$
\be\label{dJ2}
\partial_{i}H_{i+1}\equiv H_{i+1}\tJ_{i}.
\ee
The integrated Bianchi identity $H_{1}H_{2}H_{3}=1$ allows us to express the third type of $H$ derivatives in terms of these currents
\be
\label{dJ3}
\partial_{i}H_{i-1}= -H_{i-1}J_{i}-  \tJ_{i}H_{i-1}.
\ee
Finally, derivatives of the currents $\tJ$ can be purely expressed in terms of $J$ derivatives:
\be\label{ddJ1}
\partial_{i+1}\tJ_{i}= H_{i+1}^{-1} \partial_{i}J_{i+1}H_{i+1} = H_{i+1}^{-1} B_{i-1}H_{i+1},
\ee
and 
\be\label{ddJ2}
D_{i-1}\tJ_{i}=  H_{i-1} \partial_{i-1}J_{i}H_{i-1}^{-1}-\partial_{i}J_{i-1}
\ee

The computation of (\ref{main}) splits in two parts, first one needs to evaluate 
\bes
I_{i}(y)&\equiv&  \left[
                 \partial_{i-1}^y \partial_{i-1}^z\tr_{ad}\left(H_{i}(y)(\partial_{i}^y \partial_{i}^z\Theta_{i}^\mu(y,z))(H_{i}(y))^{-1}\right) 
                 \right]_{y=z}\\
&=& \int \extd x \tr_{ad}\left(\left[ \partial_{i-1}^y\left(H_{i}(y)(\partial_{i}^y {\bf{G}}_{i}^\mu(y,x)\right)\right]
  \left[ \partial_{i-1}^y\left(\partial_{i}^y {\bf{G}}_{i}^\mu(x,y)\right)H_{i}^{-1}(z)\right] \right) .\nonumber
\ees
One starts by the  evaluation of 
\bes\label{G1}
& &\partial_{i-1}^y\left(H_{i}(y)(\partial_{i}^y {\bf{G}}_{i}^\mu(y,x)\right)
 \\
&=& (\partial_{i-1}^y H_{i}(y)\Lambda_{i}(y,x)) \delta_{\mu}(y,x) + H_{i}(y)\Lambda_{i}(y,x)\partial_{i-1}^y\delta_{\mu}(y,x)
\nonumber
\ees
where we have used the property (\ref{lprop}) $\partial_{i}^y \Lambda_{i}(y,x) =0$. Moreover, choosing $i=1$ without loss of generality 
and starting from the definitions  of $\Lambda_{1}(y,x) $ and the currents one can compute
\bes\label{HL1}
&&\partial_{3}^y  H_{1}(y)\Lambda_{1}(y,x)=
H_{1}(y) \left[\tJ_{3}(y_{1},y_{2},y_{3})-\tJ_{3}(x_{1},y_{2},y_{3}) \right. \\  \nonumber
&& +\left. H_{2}(x_{1},y_{2},y_{3})\left(J_{3}(x_{1},x_{2},y_{3})-J_{3}(x_{1},y_{2},y_{3})\right)H_{2}^{-1}(x_{1},y_{2},y_{3})\right]\Lambda_{1}(y,x).
\ees
We can Taylor expand 
(\ref{G1})  and get
\bes\label{G1x}
& &\partial_{3}^y\left(H_{1}(y)(\partial_{1}^y {\bf{G}}_{1}^\mu(y,x)\right)
 \\
&=&  \left[-X_{1}B_{2}(y)+ X_{2}(H_{3}^{-1}B_{1}H_{3})(y)+ 2\mu^2 X_{3}\right]\delta_{\mu}(X) H_{1}(y)+ O(X^2)\nonumber
\ees
with $X_{i}\equiv x_{i}-y_{i} $ and
where we have made use of the relations (\ref{dJ2}), $\Lambda(x,x)=1$ and 
the explicit form  of the regulated delta function 
derivative:  $\partial^{y}_{3}\delta_{\mu}(x,y)= 2\mu^{2}(x_{3}-y_{3})\delta_{\mu}(x,y)$.

One continues with the  evaluation of 
\bes\label{G2}
& &\partial_{i-1}^y\left((\partial_{i}^y {\bf{G}}_{i}^\mu(x,y)) H_{i}^{-1}(y)\right)
 \\
&=&-\Lambda_{i}(x,y)H_{i}^{-1}(y) \partial_{i-1}^{y}\delta^\mu(x,y)+  \partial_{i-1}^{y}\left(\partial_{i}^y\Lambda_{i}(x,y))H_{i}^{-1}(y) G_{i}^\mu(x,y)\right)
\nonumber
\ees
with 
\bes\label{HL2}\nonumber
&&\partial_{1}^y \Lambda_{1}(x,y)=\Lambda_{1}(x,y)H_{1}^{-1}(y)\left[J_{1}(y_{1},y_{2},y_{3})-J_{1}(y_{1},y_{2},x_{3})\right]H_{1}(y)
\\  \nonumber
&& +H_{2}(y_{1},x_{2},x_{3}) \left[\tJ_{1}(y_{1},x_{2},x_{3})-\tJ_{1}(y_{1},y_{2},x_{3})
\right]H_{2}^{-1}(y_{1},x_{2},x_{3}) \Lambda_{1}(x,y),
\ees
and
\be
\partial_{3}^y (\partial_{1}^y \Lambda_{1}(x,y)) H_{1}^{-1}(y)= \Lambda_{1}(x,y)H_{1}^{-1}(y) \partial_{3}J_{1}(y) .
\ee
We have also used the property
\be
\partial_{3}^y \Lambda_{1}(x,y) H_{1}^{-1}(y)= 0.
\ee
The Taylor expansion of (\ref{G2}) is therefore given by
\bes\label{G2x}
& &\partial_{i-1}^y\left(H_{i}(y)(\partial_{i}^y {\bf{G}}_{i}^\mu(x,y)\right)
 \\
&=& H_{1}^{-1}(y)\left [-  2\mu^2 X_{3}\delta_{\mu}(X)+B_{2}(y)G_{1}^\mu(X)\right.\nonumber\\
& &\left. + [X_{3}B_{2}-X_{2}(H_{1}B_{3}H_{1}^{-1})](y)\partial_{3}G_{1}^\mu(X) \right] + O(X^2)\nonumber
\ees
we have denoted $G_{i}^\mu(X)\equiv G_{i}^\mu(x,y)$ and similarly for $\delta_{\mu}$. 

Putting these results together one  obtains
\bes
I_{1}= \int \extd X \tr_{ad} \nonumber
\left( 
                 \left[-X_{1}B_{2}(y)+ X_{2}(H_{3}^{-1}B_{1}H_{3})(y)+ 2\mu^2 X_{3}\right]\delta_{\mu}(X) \right. \\
\left. \left [-  2\mu^2 X_{3}\delta_{\mu}(X)+B_{2}(y)(1-2\mu^2X_{3}^2)G_{1}^\mu(X)+ 2\mu^2X_{2}X_{3}(H_{1}B_{3}H_{1}^{-1})(y)G_{1}^\mu(X) \right]\right)\nonumber
\ees
where we have neglected inside the parenthesis in the integrand all the term of cubic and higher orders in $X_{i}$ and
express $\partial_{3}G_{1}=-2\mu^2X_{3}G_{1}$. 
Due to  parity symmetry only the terms containing $X_{i}^{2}\delta_{\mu}^2(X)$, $X_{i}G_{i}\delta_{\mu}$ or $X_{j}^2X_{i}G_{i}\delta_{\mu}$ 
 are non zero, the corresponding integrals can be explicitly evaluated
\bes
\int \extd X X_{1}^2\delta_{\mu}^2(X)& =& \frac{\mu}{4(2\pi)^{\frac32}},\\
\int \extd X X_{1}G_{1}(X)\delta_{\mu}(X) &=& 2\frac{\mu}{4(2\pi)^{\frac32}},\\
2\mu^2\int \extd X X_{3}^2 X_{1}G_{1}(X)\delta_{\mu}(X) & = &\frac{\mu}{4(2\pi)^{\frac32}},\\
2\mu^2\int \extd X X_{3}^2 G_{1}(X)G_{1}(-X)& = &2 \frac{\mu}{4(2\pi)^{\frac32}}.
\ees
(the last integral will be used later.)
Moreover for the same reason the cubic terms which have been neglected do not contribute to the integral, the first non trivial contribution starts at 
quartic order and the corresponding integral is of order $1/\mu$, which justifies,  a posteriori, our approximation. 
Eventually, using the relation $\tr_{ad}(B^2)= 2 N\tr(B^2) $,
$\tr_{ad}(1)=N^2-1$ one gets 
\be
I_{1}= \frac{\mu }{4(2\pi)^{\frac32}}\left( -2N\tr(B_{2}^2 ) - 4\mu^4 ({N^2-1})\right) +O\left(\frac1\mu\right)
\ee
which proves (\ref{id1}).

For the second part of the computation one needs to evaluate
\bes\label{2I}
II_{i}(y)&\equiv&  \left[
\partial_{i-1}^y \partial_{i-1}^z 
           \tr_{ad} \left(
                            D_{i}^y\bar{D}_{i}^{z} \Theta_{i-1}^\mu(y,z) 
                            \right)
\right]_{y=z}\\
&=& \int \extd x \tr_{ad}\left([\partial_{i-1}^{y}D_{i}^y{\bf G}_{i-1}^{\mu}(y,x) ] 
[\partial_{i-1}^{y}\bar{D}_{i}^y{\bf G}_{i-1}^{\mu}(x,y)] \right)\nonumber
\ees
One first focus on the derivative (take $i=2$)
\bes \label{G3}
& &[\partial_{1}^{y}D_{2}^y{\bf G}_{1}^{\mu}(y,x) ] \\
&=& (\partial_{1}^{y}D_{2}^y\Lambda_{1}(y,x)) G_{1}^{\mu}(y,x) + (D_{2}^y\Lambda_{1}(y,x)) \delta_{\mu}(y,x) + \Lambda_{1}(y,x) \partial_{2}^y\delta_{\mu}(y,x),\nonumber
\ees
with
\bes\label{DG3}
D_{2}^{y}\Lambda_{1}(y,x) &=& H_{2}(y)(\partial_{2}^{y}H_{2}^{-1}(y)\Lambda_{1}(y,x))\\
 &=& (J_{2}(x_{1},y_{2},y_{3})- J_{2}(y_{1},y_{2},y_{3}))\Lambda_{1}(y,x)
\ees
and 
\be
\partial_{1}^{y}D_{2}^y\Lambda_{1}(y,x) = -\partial_{1}J_{2}(y) \Lambda_{1}(y,x)= -B_{3}(y) \Lambda_{1}(y,x).
\ee
The equality (\ref{DG3}) follows from the definition of $D_{i}$ and the next one from a direct computation.
The Taylor expansion of the first term is therefore given by
\bes \label{G3x}
& &[\partial_{1}^{y}\bar{D}_{2}^y{\bf G}_{1}^{\mu}(y,x) ] \\
&=& (B_{3}(y)(X_{1}\delta_{\mu}(X)-G_{1}(-X))+ 2\mu^{2}X_{2}\delta_{\mu}(X))+ O(X^{2})
\ees
with $X_{i}=x_{i}-y_{i}$.

The second term of (\ref{2I}) is given by 
\bes \label{G4}
[\partial_{1}^{y}\bar{D}_{2}^y{\bf G}_{1}^{\mu}(x,y) ]
&=& (\partial_{1}^{y}\bar{D}_{2}^y\Lambda_{1}(x,y)) G_{1}^{\mu}(x,y) - (\bar{D}_{2}^y\Lambda_{1}(x,y)) \delta_{\mu}(x,y) \nonumber \\
& & + \nonumber
(\partial_{1}^{y}\Lambda_{1}(x,y)) \partial_{2}^{y}G_{1}(x,y) - \Lambda_{1}(x,y) \partial_{2}^y\delta_{\mu}(x,y),\nonumber
\ees
with
\bes
\bar{D}_{2}\Lambda_{1}(x,y) &=&\partial_{2}^{y}(\Lambda_{1}H_{2})(x,y)H_{2}^{-1}(y) \\
\nonumber &=& H_{2}(y_{1},x_{2},x_{3})H_{3}(y_{1}, y_{2},y_{3})[\tJ_{2}(y_{1}, y_{2},x_{3})- \tJ_{2}(y)]H_{1}(y)\\
&=& X_{3} (H_{2}B_{1}H_{2}^{-{1}})(y) + O(X^{2})
\ees
and 
\bes
\partial_{1}^{y}\bar{D}_{2}\Lambda_{1}(x,y)&=& X_{3}[ H_{2}(\tilde{D}_{1}B_{1})H_{2}^{-{1}}](y) + O(X^2),\\
\partial_{1}\Lambda_{1}(x,y)&=& X_{2}B_{3}(y)-X_{3}(H_{1}^{-1}B_{2}H_{1})(y) + O(X^2),
\ees
with $\tilde{D}_{i}= \partial_{i}+ \tJ_{i}$.
Thus, up to terms of order $1/\mu$ we have 
\bes
II_{2}(y)&=& \int \extd X \tr_{ad}\left( \left[B_{3}(y)(X_{1}\delta_{\mu}(X)-G_{1}(-X))+ 2\mu^{2}X_{2}\delta_{\mu}(X)\right] \right.\\
 & &\times\nonumber\left[X_{3}\{H_{2}\tilde{D}_{1}B_{1}H_{2}^{-1}G_{1}^{\mu}(X)-H_{2}B_{1}H_{2}^{-1}\delta_{\mu}(X)-H_{1}B_{2}H_{1} 2\mu^{2}G_{1}^{\mu}(X)\}\right.\\
& & \quad \nonumber\left. \left. + B_{3}\, 2\mu^2 X_{2}^{2}G_{1}^{\mu}(X) - 2\mu^{2}X_{2}\delta_{\mu}(X)\right]^{}\right)
\ees
Due to parity symmetry the term proportional to $X_{3}$ do not contribute, we are left with
\bes\nonumber
II_{2}(y)&=& \tr_{ad}(B_{3}^{2}(y))  \int \extd X X_{2}^{2}G_{1}(X)(X_{1}\delta_{\mu}(X)-G_{1}(-X))
\\ & & \quad -4\mu^{4}\tr_{ad}(1) \int \extd X X_{2}^{2} \delta_{\mu}(X)^{2}\nonumber\\
&=&  \frac{\mu }{4(2\pi)^{\frac32}}\left( -2N\tr(B_{3}^{2}(y)) - 4\mu^{4}(N^{2}-1)\right).\nonumber
\ees
This proves (\ref{id2}).


\begin{thebibliography}{100}


\bibitem{KN}
D.~Karabali and V.~P.~Nair,
{\em ``A gauge-invariant Hamiltonian analysis for non-Abelian gauge theories in
(2+1) dimensions,''}
Nucl.\ Phys.\ B {\bf 464}, 135 (1996);\\
D.~Karabali and V.~P.~Nair,
 {\em ``On the origin of the mass gap for non-Abelian gauge theories in (2+1)
dimensions,''}
Phys.\ Lett.\ B {\bf 379}, 141 (1996);

\bibitem{KKN}
D.~Karabali, C.~j.~Kim and V.~P.~Nair,
{\em ``Planar Yang-Mills theory: Hamiltonian, regulators and mass gap,''}
Nucl.\ Phys.\ B {\bf 524}, 661 (1998)
[arXiv:hep-th/9705087];\\
D.~Karabali, C.~J.~Kim and V.~P.~Nair,
{\em ``On the vacuum wave function and string tension of Yang-Mills theories  in
(2+1) dimensions,''}
Phys.\ Lett.\ B {\bf 434}, 103 (1998)
[arXiv:hep-th/9804132];\\
D.~Karabali, C.~J.~Kim and V.~P.~Nair,
{\em ``Manifest covariance and the Hamiltonian approach to mass gap in  (2+1)-
dimensional Yang-Mills theory,''}
Phys.\ Rev.\ D {\bf 64}, 025011 (2001)
[arXiv:hep-th/0007188] and references therein.

\bibitem{Schulz}
  H.~Schulz,
  ``The 3-D Yang-Mills system,''
  arXiv:hep-ph/0008239.
\bibitem{Feynman}
  R.~P.~Feynman,
  ``The Qualitative Behavior Of Yang-Mills Theory In (2+1)-Dimensions,''
  Nucl.\ Phys.\ B {\bf 188}, 479 (1981).

\bibitem{prl}
 R.~G.~Leigh, D.~Minic and A.~Yelnikov,
  ``Solving pure QCD in 2+1 dimensions,''
  arXiv:hep-th/0512111.

\bibitem{long}
R.~G.~Leigh, D.~Minic and A.~Yelnikov,
  ``How to solve pure yang-Mills theory in 2+1 dimensions,''
  arXiv:hep-th/0604060.
 

\bibitem{Teper}
M.~J.~Teper,
  ``SU(N) gauge theories in 2+1 dimensions,''
  Phys.\ Rev.\ D {\bf 59}, 014512 (1999)
  [arXiv:hep-lat/9804008].

\bibitem{jointpaper}
L.~Freidel, R.~G.~Leigh, D.~Minic, ``Towards a solution of pure Yang-Mills theory
in $3+1$ dimensions''.

\bibitem{bars}
I.~Bars, Phys.\ Rev.\ Lett. {\bf 40} 688 (1978) ; Nucl.\ Phys. \ B {\bf 149} 39 (1979);
also, I. Bars and F. Green, Nucl.\ Phys. \ B {\bf 148} 445 (1979), Erratum-ibid. B {\bf 155} 543 (1979).
The large N limit was explored in
I.~Bars, Phys. Lett. \ B {\bf 116} 57 (1982); Phys.\ Lett.\ B {\bf 245} 35 (1990).


\bibitem{GK}
 K.~Gawedzki and A.~Kupiainen,
  ``G/H Conformal Field Theory From Gauged WZW Model,''
  Phys.\ Lett.\ B {\bf 215}, 119 (1988).





\bibitem{NAlex}
  R.~Efraty and V.~P.~Nair,
  ``Chern-Simons theory and the quark - gluon plasma,''
  Phys.\ Rev.\ D {\bf 47}, 5601 (1993)
  [arXiv:hep-th/9212068]
  G.~Alexanian and V.~P.~Nair,
  ``A Selfconsistent inclusion of magnetic screening for the quark - gluon
  plasma,''
  Phys.\ Lett.\ B {\bf 352}, 435 (1995)
  [arXiv:hep-ph/9504256].


\bibitem{KS}
  J.~B.~Kogut and L.~Susskind,
  ``Hamiltonian Formulation Of Wilson's Lattice Gauge Theories,''
  Phys.\ Rev.\ D {\bf 11}, 395 (1975).

\bibitem{clusterexp}
J.~P.~Greensite,
  ``Large Scale Vacuum Structure And New Calculational Techniques In Lattice
  SU(N) Gauge Theory,''
  Nucl.\ Phys.\ B {\bf 166}, 113 (1980).\\
  A.~C.~Irving and C.~J.~Hamer,
  ``Methods In Hamiltonian Lattice Field Theory. 2. Linked Cluster
  Expansions,''
  Nucl.\ Phys.\ B {\bf 230}, 361 (1984).

\bibitem{dimred}
J.~P.~Greensite,
  ``Calculation Of The Yang-Mills Vacuum Wave Functional,''
  Nucl.\ Phys.\ B {\bf 158}, 469 (1979).\\
P.~Mansfield,
  ``Continuum strong coupling expansion of Yang-Mills theory: Quark confinement
  and infrared slavery,''
  Nucl.\ Phys.\ B {\bf 418}, 113 (1994)
  [arXiv:hep-th/9308116].
\\
 S.~Samuel,
  ``On the 0++ glueball mass,''
  Phys.\ Rev.\ D {\bf 55}, 4189 (1997)
  [arXiv:hep-ph/9604405].


\bibitem{nality}
  J.~Greensite and M.~B.~Halpern,
  ``Suppression Of Color Screening At Large N,''
  Phys.\ Rev.\ D {\bf 27}, 2545 (1983). 
\\
J.~Greensite and S.~Olejnik,
  ``K-string tensions and center vortices at large N,''
  JHEP {\bf 0209}, 039 (2002)
  [arXiv:hep-lat/0209088]. 
\\
 A.~Armoni and M.~Shifman,
  ``Remarks on stable and quasi-stable k-strings at large N,''
  Nucl.\ Phys.\ B {\bf 671}, 67 (2003)
  [arXiv:hep-th/0307020].


\bibitem{UV}
  D.~J.~Gross and F.~Wilczek,
  ``Ultraviolet Behavior Of Non-Abelian Gauge Theories,''
  Phys.\ Rev.\ Lett.\  {\bf 30}, 1343 (1973).
  H.~D.~Politzer,
  ``Reliable Perturbative Results For Strong Interactions?,''
  Phys.\ Rev.\ Lett.\  {\bf 30}, 1346 (1973).
  
\end{thebibliography}
\end{document}